\documentclass[aps,prb, reprint, showpacs, floatfix]{revtex4-1}
\usepackage{amsmath}
\usepackage{amssymb}
\usepackage{graphicx}
\usepackage{xcolor}

\newcommand{\figref}[1]{Fig.~\ref{#1}}
\newcommand{\eqnref}[1]{Eq.~\eqref{#1}}

\newcommand{\ket}[1]{\ensuremath{|{#1}\rangle}}

\begin{document}


\title{Dual-channel lock-in magnetometer with a single spin in diamond}
\author{N. M. Nusran}
\author{M. V. Gurudev Dutt}
\email[Correspondence to:]{gdutt@pitt.edu}
\affiliation{Department of Physics and Astronomy, University of Pittsburgh, Pittsburgh, PA 15260}	

\date{\today}

\begin{abstract}
We present an experimental method to perform dual-channel lock-in magnetometry of time-dependent magnetic fields using a single spin associated with a nitrogen-vacancy (NV) color center in diamond. We incorporate multi-pulse quantum sensing sequences with phase estimation algorithms to achieve linearized field readout and constant, nearly decoherence-limited sensitivity over a wide dynamic range. Furthermore, we demonstrate unambiguous reconstruction of the amplitude and phase of the magnetic field. We show that our technique can be applied to measure random phase jumps in the magnetic field, as well as phase-sensitive readout of the frequency.  
\end{abstract}
\maketitle

The coherent evolution of a quantum state interacting with its environment is the basis for understanding fundamental issues of open quantum systems~\cite{Zurek03}, as well as for applications in quantum information science and technology~\cite{Quantbook}. Traditionally in these fields, the extreme sensitivity of coherent quantum dynamics to external perturbations has been viewed as a barrier to be surmounted. By contrast, quantum sensors have emerged that instead take advantage of this sensitivity; recent examples include electrometers and magnetometers based on superconducting qubits~\cite{Bylander11}, quantum dots~\cite{Vamivakas11}, spins in diamond~\cite{Maze08,Taylor08,Balasub08,Dolde11} and trapped ions~\cite{Kotler11}. 
 
\begin{figure}[hbt]
\scalebox{1}[1]{\includegraphics[width=3.4in]{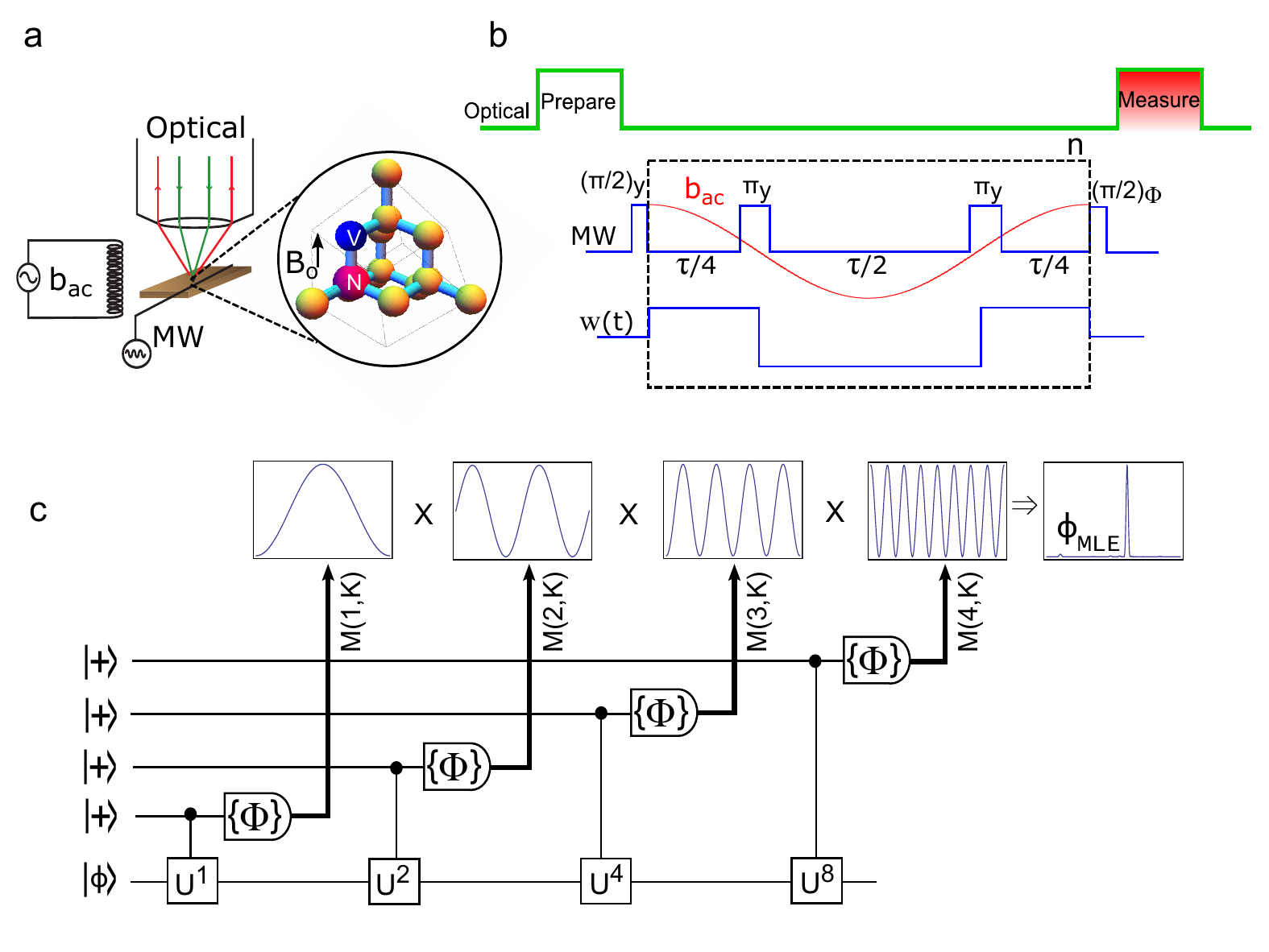}}
\caption{\small{
(a) Illustration of experimental setup for NV magnetometry (b) Carr-Purcell (CP) sequences with even number $2 n$ ($n=1,2, \ldots$) of $\pi$ pulses, are used to detect oscillating magnetic fields $b_z(t) = b_{ac} \cos (2 \pi f_{ac} t)$ where $f_{ac}$ is the AC field frequency. Here, $\pi/2$ ($\pi$) microwave pulses cause corresponding rotations of the spin vector, $\Phi$ is an adjustable control phase and $w(t)$ represents the CP filter function with reference frequency $\omega_0 = 2 \pi /\tau$. (c) Schematic quantum circuit for the phase estimation algorithm (PEA). Here $\ket{+} = (\ket{0} + \ket{1})/\sqrt{2}$ and $U^{n}$ is the unitary evolution under the action of an external drive, represented by the quantum phase $\ket{\phi}$. }}
\label{figExp}
\end{figure}

The nitrogen-vacancy (NV) defect center in diamond (\figref{figExp}(a)) shows great promise as an ultra-sensitive solid-state magnetometer and magnetic imager because it features potentially atomic-scale resolution~\cite{Taylor08}, wide temperature range operation from 4 K -- 700 K~\cite{Toyli12}, and long coherence times that allow for high magnetic field sensitivity~\cite{Balasub09}. Recent demonstrations include nanoscale magnetic imaging~\cite{Balasub08,Maletinsky12,Grinolds13}, coupling to nano-mechanical oscillators~\cite{Hong12,Arcizet11,Kolkowitz12b}, detection of single proximal nuclear spins~\cite{Dutt07,Zhao12,Taminiau12,Kolkowitz12} and nanoscale volumes of external electron and nuclear spins~\cite{Mamin12,Laraoui12,Mamin13,Staudacher13}. 

Magnetometry with diamond spin sensors detects the frequency shift of the NV spin resonance caused by the magnetic field via the Zeeman effect. Highly sensitive quantum sensing techniques use multi-pulse dynamical decoupling (DD) sequences~\cite{Taylor08,Kotler11,deLange11,Naydenov11,Bylander11,Cooper13} that are tuned to the frequency of a time-dependent field. The resulting fluctuating frequency shift is rectified and integrated by the pulse sequence to yield a detectable quantum phase, while effectively filtering out low frequency noise from the environment (\figref{figExp}(b)). Another advantage of these DD sequences is that they make the magnetometer insensitive to instabilities such as drifts in temperature or applied bias magnetic field.   

However, these state of the art quantum sensing methods also have significant drawbacks: the dynamic range is limited by the quantum phase ambiguity~\cite{Higgins07,Berry09}, the sensitivity is a highly nonlinear function of field amplitude requiring prior knowledge of a working point for accurate deconvolution, and the classical phase of the field has to be carefully controlled to obtain accurate field amplitude~\cite{Taylor08}. 

In this work, we present an experimental method that incorporates the DD sequences with phase estimation algorithms (PEA) to address these fundamental problems (see \figref{figExp}(c)). Our dual-channel lock-in magnetometer has linearized field readout and nearly decoherence-limited constant sensitivity, while offering significantly greater dynamic range.  We demonstrate unambiguous reconstruction of the amplitude and phase of the magnetic field without prior knowledge of either value. Finally, we show that our technique can be applied to measure random phase jumps in the magnetic field, and to obtain phase-sensitive field frequency readout. 

As demonstrated by conventional electronic lock-in techniques, phase information is often extremely useful in measuring important physical processes such as relaxation life-times~\cite{Lakowiczbook}, spectral and spatial diffusion~\cite{Mims61,Klauder62}. Dynamic range and constant sensitivity may be helpful in accurate measurements of small magnetic fields due to spins in nanoscale volumes~\cite{Mamin13,Staudacher13}, and in measuring spin density of heterogenous samples at low magnetic fields, where different magnetic species are not well resolved in frequency space. Observables like the field phase and frequency may also be useful in situations where the target spins are hard to polarize or to drive~\cite{Rugar04,Poggio10}. Random phase and spin configuration approximations are often made to theoretically deal with this problem and thereby retrieve the field amplitude~\cite{Laraoui10,deLange11,Staudacher13,Mamin13}, but it remains to be verified if these approximations are valid experimentally in samples of interest~\cite{Degen07}. 
Thus, our methods open up the potential for new modes of magnetometry with NV spin sensors, as well as for other quantum sensors. 

The important feature of PEAs that permits this reconstruction is reminiscent of a Fourier series in the quantum phase $\phi$. Indeed, PEAs were first introduced for the purpose of performing a quantum fourier transform in Shor's algorithm. The quantum circuit representation of our PEA is shown in \figref{figExp}(c). The spin qubit is first initialized into the $\ket{+} = (\ket{0} + \ket{1})/\sqrt{2}$ state, while the auxiliary qubit is initialized into the state $\ket{\phi}$ where $\phi$ is the quantum phase to be estimated. The action of the controlled-$U^n$ gate on the auxiliary qubit register is $U^n \ket{\phi} = e^{i n \phi} \ket{\phi}$, resulting in the combined state 
\[ \frac{\ket{0} + e^{i n \phi} \ket{1}}{\sqrt{2}}\ket{\phi} \]
Measurements are carried out in the set of basis states $\{ \ket{\Phi} \}$ and the resulting probability distributions are combined using Bayesian analysis to obtain an estimate for the state \ket{\phi}. Since the auxiliary registers are not measured, they can be replaced by a classical drive field that causes the phase shift shown above on the control qubit, and the corresponding quantum phase allows us to estimate the classical drive field. Our implementation of the PEA is discussed below, after we introduce our experimental system and quantum sensing with DD sequences.

The NV center is a spin-1 system in the ground state, quantized along the C$_{3v}$ symmetry axis, with the $\ket{m_s = 0}$ and $\ket{m_s = \pm 1}$ levels split by 2.87~GHz at zero magnetic field. The spin state can be initialized by optical pumping with 532 nm laser excitation, and the spin polarization can be detected by measuring the spin-dependent fluorescence signal. Our magnetometry setup is shown schematically in \figref{figExp}(a). We use a single NV center in a type-IIa bulk diamond sample, and apply a static magnetic field $B_0$ oriented along the NV centers $z$-axis, allowing us to form a pseudo-spin $\sigma = 1/2$ qubit system with the $\ket{m_s = 0} \leftrightarrow \ket{m_s = -1}$ spin states. Microwave pulses are applied to the NV center using a thin copper wire on the diamond surface, allowing us to attain a $\pi/2$ rotation in $\sim 25$~ns. 

\begin{figure}[tbh]
\centering
\scalebox{1}[1]{\includegraphics[width=3.4in]{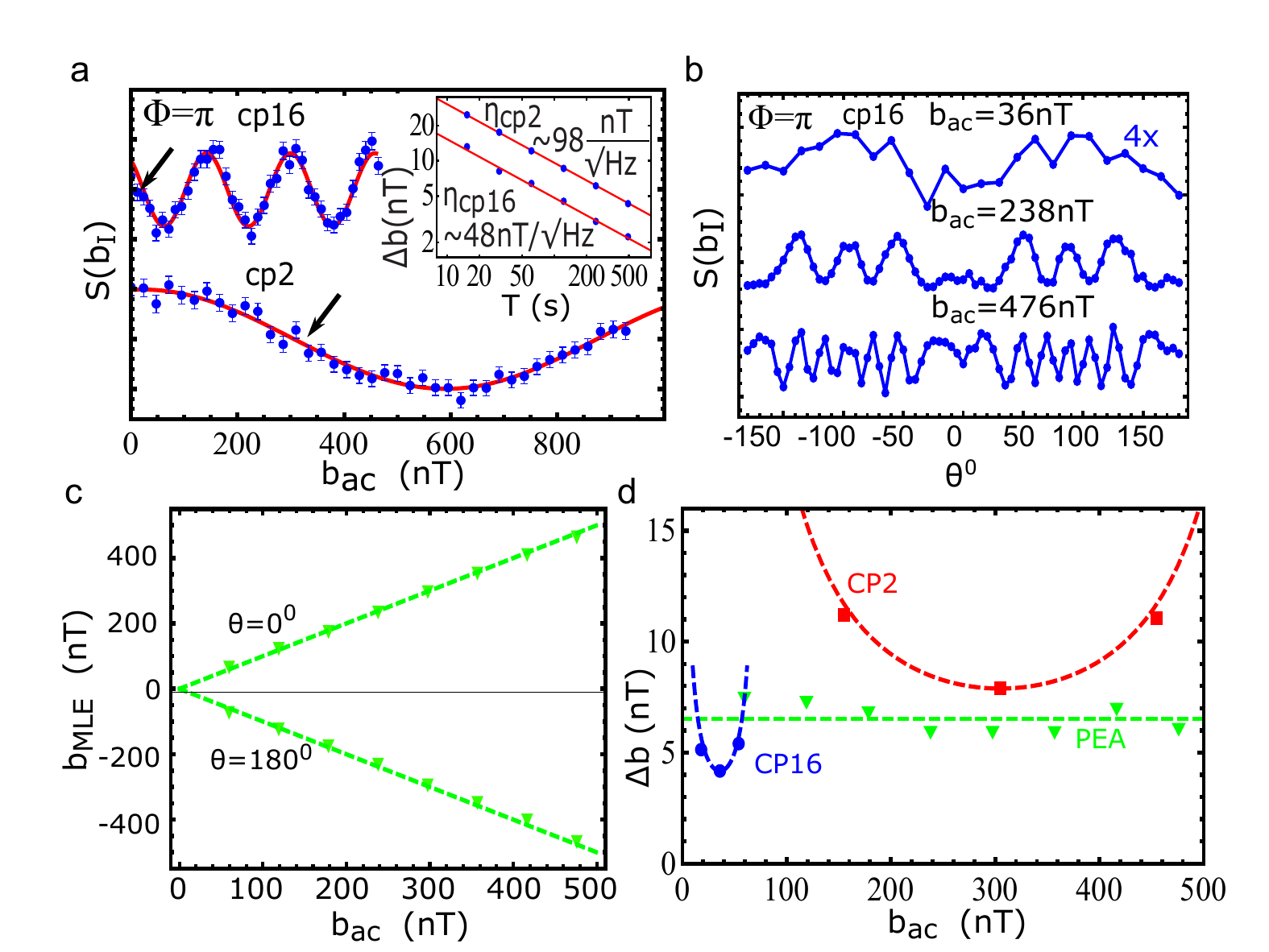}}
\caption{\small{
(a) Magnetic field dependence of signal from CP-2 (CP-16) sequences, showing the tradeoff between dynamic range and sensitivity. (b) Variation in signal as function of $\theta$ from CP-16 sequences for different magnetic fields $b_{ac}$. Data has been offset and scaled for clarity. (c) Measured value $b_{MLE}$ from PEA vs. the applied drive field $b_{ac}$ for $\theta = 0^\circ, 180^\circ$. Dashed lines represent ideal estimation $b_{MLE} = b_{ac}$. (d) Variation in the minimum detectable field $\Delta b = \eta/\sqrt{T}$ vs. $b_{ac}$ for CP sequences and PEA. Here $T = 150$~sec. Dashed lines for CP sequences are fits to \eqnref{eq:eta}. Data in this and all subsequent figures was taken with AC magnetic field frequency$f_{ac}=20.83$~kHz while the applied DC bias magnetic field is $B_{0} \approx 470$~Gauss.}}
\label{figPEA}
\end{figure} 
The Hamiltonian for a spin-$1/2$ qubit interacting with an external magnetic field $b_z (t)$, while being driven with on-resonance electromagnetic fields is given in the rotating frame by $H = \hbar \gamma_e b_z (t) \sigma_z + \hbar \Omega (\sigma_x \cos \Phi + \sigma_y \sin \Phi) $. Here $\gamma_e = 2 \pi (27.99)$~GHz/T is the gyromagnetic ratio of the spin, $\Omega$ is the Rabi frequency of the on-resonance drive field and $\Phi$ is an adjustable control phase of the microwave.  An oscillating magnetic field $b_z (t) = b_{ac} \cos (2 \pi f_{ac} t - \theta) $ can be measured by the Carr-Purcell $CP-2n$ ($n = 1, 2, \ldots$) sequence, as shown in \figref{figExp}(b). Here, $w(t)$ is the DD filter function of the CP sequence~\cite{Taylor08,Zhao12,Taminiau12,Kolkowitz12} with reference frequency $\omega_0 = 2 \pi / \tau$. The collapses and revivals in the signals due to $^{13}$C nuclear spins in our sample, shown in Supplementary Information (SI) Fig.~S1, restricts the allowable values of the filter reference frequency $\omega_0 = \omega_L/(2 p)$ where $p = 1, 2, \ldots$ is an integer, $\omega_L = \gamma_{n} B_0$ is the $^{13}$C Larmor frequency, $\gamma_{n} = 2 \pi (10.705)$~MHz/T is the $^{13}$C nuclear gyromagnetic ratio. We can then detect the in-phase magnetic field ($b_I = b_{ac} \cos \theta$) by measuring the probability $P(0)$ to be in $\ket{m_s=0}$; 
\begin{equation} 
S(b_I) = 2 P(0) - 1 = D(n\tau)  \cos (n \phi_I - \Phi) 
\label{eq:SbI}
\end{equation}
and the quantum phase 
\begin{equation}
\phi_I = \int_{0}^{ \tau} \gamma_e b_z(t) w(t) dt = 2 \gamma_e b_I \tau/\pi 
\label{eq:phI}
\end{equation}
where $D(n\tau) = \exp(- (n\tau/T_{2}^{(2n)})^3)$ is the decoherence function (see SI Fig.~S2), and $T_2^{(2n)}$ is the effective coherence time under the DD pulse sequence~\cite{Taylor08,deLange11,Naydenov11}. 

The quantum phase $\phi_I$ is ambiguous due to the multi-valued inverse sine or cosine functions, requiring us to restrict its range to $(-\pi/2n, \pi/2n)$. We also need to know the classical phase $\theta$ to obtain $b_{ac}$ or vice-versa. Furthermore, even if the classical phase is known, the dynamic range is limited by the above restriction on $\phi_I$, as shown in \figref{figPEA}(a). The sensitivity improves with higher number $n$ of $\pi$ pulses, but correspondingly only a small range of fields can be sensed.  As shown by the data in \figref{figPEA}(b), when the classical phase $\theta$ is allowed to vary, magnetic field values that differ by factors of 2 (or even 10) could yield the same signal. 

In a conventional electronic single channel lock-in amplifier, simply tuning the phase of the reference channel to minimize  or maximize the signal would allow us to find the amplitude and phase of the input signal. However, that requires a linear readout of the quantum phase $\phi_I$ which is not available directly for the CP sequences. Further, in quantum sensing, the relevant reference phase corresponds to that of the DD filter function $w(t)$, and one must adjust the timing offset of the sequence for each frequency that has to be detected. Previous works such as Refs.~\cite{Kotler11,Taylor08, Maze08, deLange11,Naydenov11} carried out this phase adjustment, usually by carefully modulating the signal AC field, to $0 (\pi/2)$ prior to measurement. Practical situations where the magnetic field arises from unknown samples may prevent this phase adjustment and result in inaccurate measurements which we address here.

Linearizing the signal could be accomplished under the assumption that both $b_{ac}$ and $\theta$ are small, and by choosing the control phase $\Phi = \pi/2$. Alternately, one can choose a working point with finite $b_{ac}$ and $\theta = 0$ (see \figref{figPEA}(a)) and look for deviations from this point. By recording and averaging the fluorescence measurements separately at these carefully adjusted working points we get the sensitivity,
\begin{equation}
\eta = \frac{1}{V \sqrt{\xi}}\frac{\pi }{2\gamma_e \sqrt{n \tau} D(n\tau)| \cos (n \phi_I)|} 
\label{eq:eta}
\end{equation}
 where $V \sim 0.3$ is the fringe visibility, and $\xi$ is a factor that depends on the photon collection efficiency in our system~\cite{Taylor08, Nusran12}. Through prior knowledge of the working point, it is assumed that $n \phi_I \approx 2 m \pi$  and this requirement will be more stringent as $n$ increases~\cite{Taylor08}. Thus, the minimum detectable field and corresponding deconvolution of the target spin positions will depend on this knowledge. Lastly, we note (and show below) that the dependence on the frequency of $b_z(t)$ for the CP pulse sequences is symmetric around the lock-in reference frequency $\omega_0 = 2\pi/\tau$, implying that frequency changes of the field are also ambiguous. 

Overcoming the multiple ambiguities of the quantum phase caused by uncertainty in the parameters of the external magnetic field is thus an important step. Recently, phase estimation algorithms (PEAs) were introduced for DC magnetic field sensing with single spins in diamond~\cite{Said11, Nusran12,Waldherr12}. We first extend these results and demonstrate significant improvement in the dynamic range and linearization of the field readout, by combining the DD pulse sequence ($CP - 2n$) with PEA. 

In our application, the DD pulse sequence times are first increased in powers of 2 (i.e. $n = 2^{k-1}$ with $k = 1, \ldots, K$) starting with an initial time that is determined by the fundamental reference frequency ($\omega_0$) that one wishes to sense. This results in phase accumulations $\phi_k = 2^{k - 1}\phi$ at each step of the PEA, as shown in \figref{figExp}(c). Secondly, measurements with the smallest times (low Zeeman shift resolution) are corrected by repeating them several times, analogous to the weighting coefficient in a Fourier series with the weighting factors $M(K,k)= M_K + F (K-k)$, with $M_K, F$ optimized through numerical simulations (see SI and Refs.~\cite{Berry09,Said11}). Finally, the control phase of the readout pulse is cycled through several values to measure along different basis vectors, thus allowing to differentiate between quantum phases that differ by fractions of $\pi$. These steps are combined with digitization of the signal levels, and Bayesian maximum likelihood filtering to obtain unambiguous knowledge of the field value. See SI Fig.~S3 for description of the Bayesian estimation process~\cite{Said11,Nusran12,Waldherr12}.

The data in \figref{figPEA}(c) shows that our PEA unambiguously measures the value of the magnetic field with a linear readout over a wide range, and is also able to resolve phase shifts of $\pi$. 
We now turn to the comparison of the minimum detectable field $\Delta b= \eta/ \sqrt{T}$ obtained in both approaches for some fixed averaging time $T$, where we chose $T = 150$~sec as a typical time used in sensitive experiments~\cite{Grinolds13,Mamin13,Staudacher13}. As expected from \eqnref{eq:eta}, the CP sequences show rapid degradation in $\Delta b $ as soon as we  deviate from the working point, for instance due to imperfect knowledge of $\theta$. By contrast, \figref{figPEA}(d) shows that the sensitivity achieved by PEA remains almost a constant over a wide range of $b_{ac}$ and is comparable to the longest CP sequence used in our work. The maximum detectable field ($\pm b_{ac,max}$) of the PEA is obtained by setting $\phi_I = \pm\pi$ in \eqnref{eq:phI}, 
\begin{equation}
b_{ac,max} =  \frac{\pi \omega_0}{4 \gamma_e}
\end{equation}
which in principle has no fundamental limit except for the restriction $\omega_0 = \omega_L/(2 p)$ mentioned previously for our samples. The dynamic range (DR) is given by,
\begin{equation}
DR = \frac{b_{ac,max}}{\Delta b}
\end{equation}
and from the data in \figref{figPEA}(c), we obtain $DR^{CP-16} \sim 3.4$, while $DR^{PEA} \sim 90$. As shown in the SI (Fig.~S5), $DR^{PEA}$ keeps increasing at higher frequencies, while by contrast, the $DR^{CP}$ is essentially unchanged. 

\begin{figure}[tbh]
\scalebox{1}[1]{\includegraphics[width=3.4in]{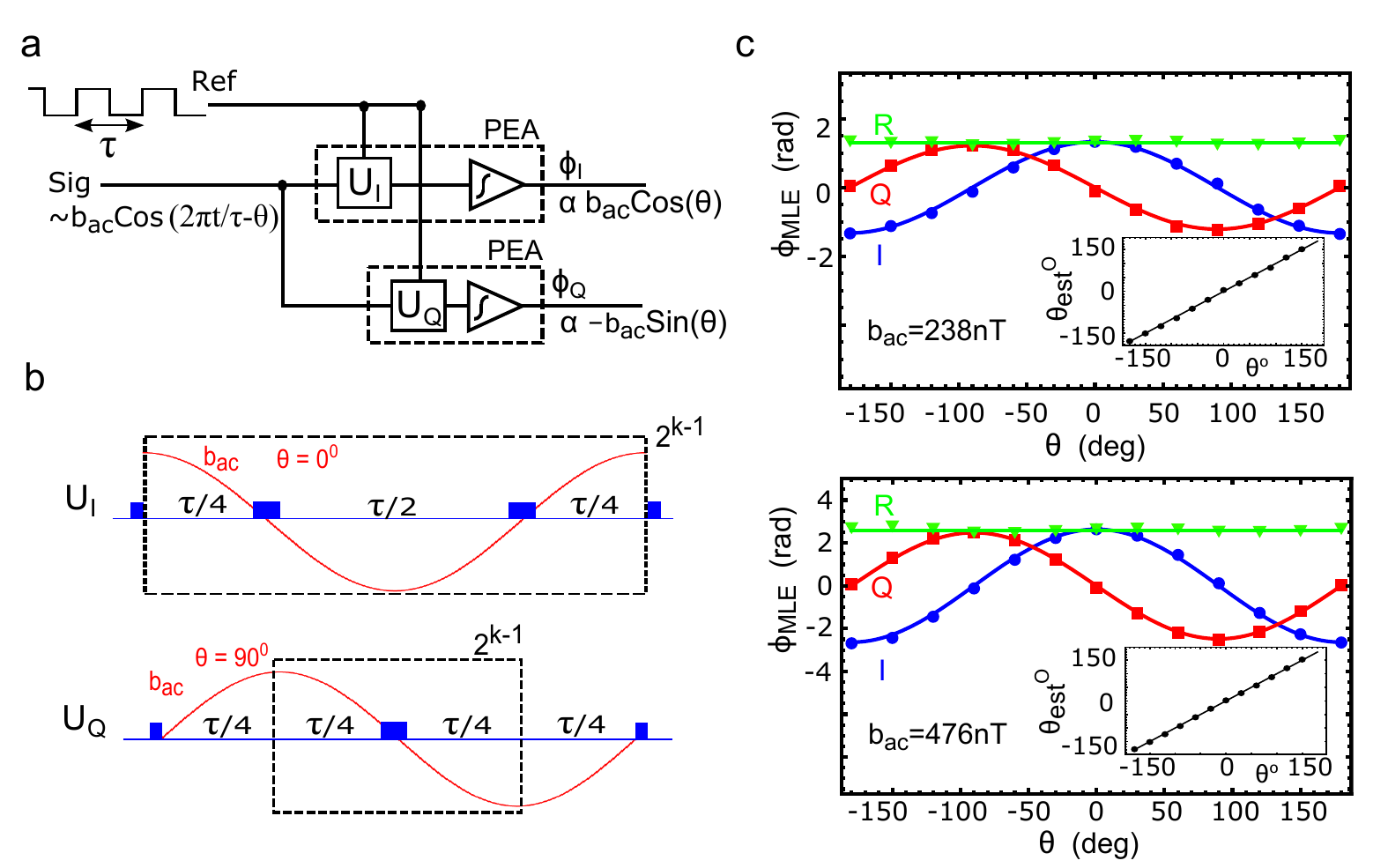}}
\caption{\small{
(a) Schematic illustration of the quantum dual-channel lock-in magnetometer. Via unitary evolution of the single spin, the applied magnetic field is multiplied with the lock-in reference signal set by the DD pulse sequences. The PEA is implemented for each channel as before to linearize the readout and yield the I and Q quantum phases.(b) DD pulse sequences for $U_I$ and $U_Q$. (c), (d) Data for $\phi_I$ and $\phi_Q$  as $\theta$ is varied for different values of $b_{ac} = 238 (476)$~nT. The estimator $\phi_R$ remains constant throughout, solid lines represent ideal sensing. (inset) Data for $\theta_{est}$ as function of $\theta$, solid line represents ideal case $\theta_{est} = \theta$.}}
\label{figLockin}
\end{figure}

 Our dual-channel quantum lock-in magnetometer scheme, shown schematically in \figref{figLockin}(a), detects both in-phase $b_I = b_{ac} \cos \theta $ and quadrature components $b_Q = - b_{ac} \sin \theta $ of the magnetic field. As noted above, using CP sequences alone does not allow us to obtain both components unless we have excellent knowledge of both $b_{ac}$ and $\theta$. We further modify the PEA algorithm for lock-in detection by using both the CP-$(2n-1)$ and CP-$2n$ pulse sequence depicted in \figref{figLockin}(b) to obtain unambiguous information about the magnetic field quadratures. The former case is sensitive to $\theta = \pm 90^\circ$, whereas the latter case  is sensitive to $\theta = 0 (180)^\circ$. As shown in \figref{figLockin}(c), we can determine for various $b_{ac}$, the estimators 
\begin{align}
\theta_{est} &= \tan^{-1} (-\phi_Q/\phi_I) \\
\phi_R &= \sqrt{\phi_I^2 + \phi_Q^2}
\end{align}
where $\phi_{Q} = 2 \gamma_e b_Q \tau/ \pi$ and thus reconstruct $b_{ac}^{MLE} = \tfrac{\pi \phi_R}{2 \gamma_e \tau}$ independent of the value of $\theta$. 

The phase resolution of our lock-in magnetometer is given by the sample standard error of our estimator $\theta^{est}$ from the actual value $\theta$ used in the experiment,
\begin{equation}
\Delta \theta_{min} = \sqrt{\frac{1}{N(N - 1)} \sum_{i=1}^N (\theta_{est,i} - \theta)^2} 
\end{equation}
which evaluates to $\Delta \theta_{min} = 0.3^\circ (0.6^\circ)$ for $b_{ac} = 476 (238)$~nT. For ease of data analysis in our experiments, we carried out $I$ and $Q$ PEA routines successively, however this can easily be modified to have both sequences alternated within one PEA routine for near-simultaneous detection of the quadratures.

\begin{figure}[hbt]
\scalebox{1}[1]{\includegraphics[width=3.4in]{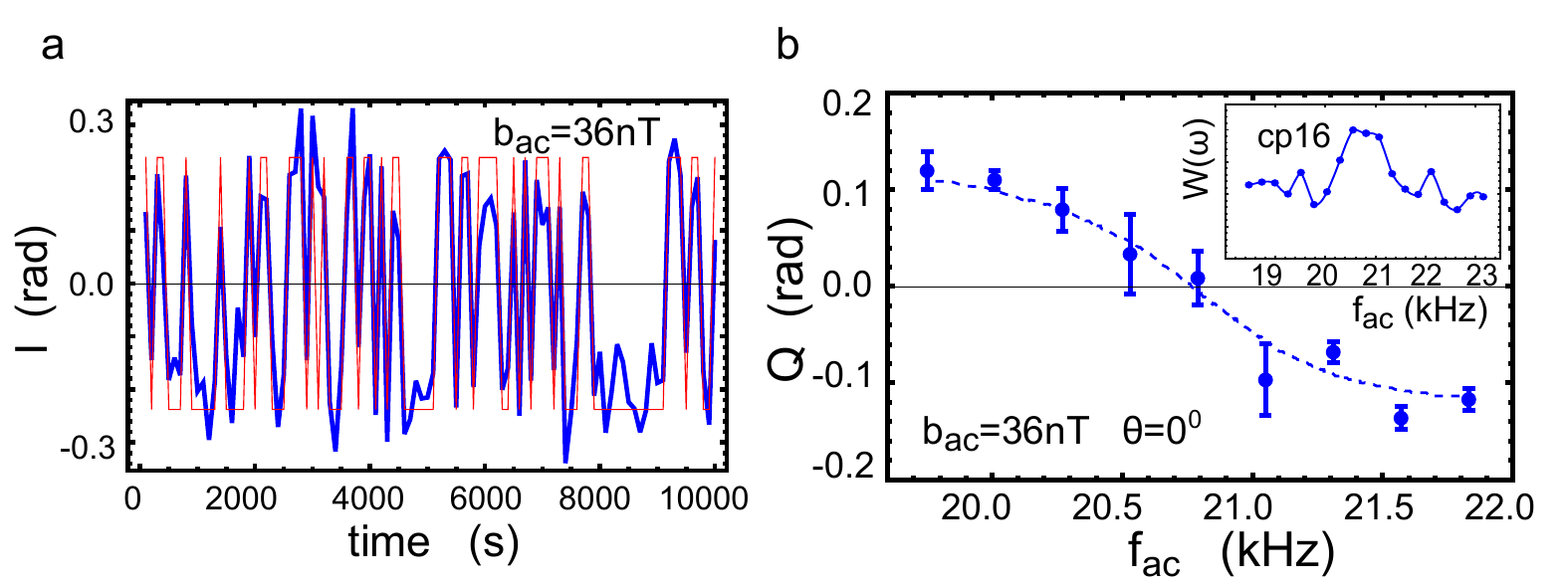}}
\caption{\small{
(a) Observation of random telegraph phase jumps $0 \leftrightarrow \pi$ by monitoring the $I$ channel of the lock-in. The time constant of the lock-in was set to $\sim 100$~sec, and the phase was held constant over this time but allowed to vary randomly between measurements. Thick (blue) lines represent observed phase, and thin (red) lines represent actual phase. (b) Phase sensitive readout of frequency change in the magnetic field. (inset) Frequency response of CP-16 sequence.}}
\label{figFreq}
\end{figure}

We now demonstrate two important applications of our dual-channel lock-in magnetometer. Earlier work has studied the effect of random classical phase on the magnetometry performance of DD sequences~\cite{Laraoui10,deLange11}. However, these methods require theoretical assumptions on the nature of the random phase e.g. uniformly or normally distributed. In \figref{figFreq}(a), we show that by monitoring the $\phi_I$ channel of our lock-in, we can observe random telegraph phase flips $0 \leftrightarrow \pi$ of the magnetic field. One physical scenario where such jumps might occur in the phase would be for measurements of single electron or nuclear spins where the spins cannot be easily polarized, but will be present in one state or the other for each measurement shot. Similarly, when nanoscale volumes of spins are measured experimentally~\cite{Mamin13,Staudacher13}, one could use this protocol to verify that we sample all possible spin configurations by either periodically randomizing the ensemble or simply by waiting for long enough durations~\cite{Degen07}.  See SI Fig.~S4 for data similar to such situations.

Our second application is for measurement of the frequency changes from the central working frequency $\omega_0 = 2 \pi/\tau$. As shown in the inset to \figref{figFreq}(b), the in-phase response of the CP sequences is symmetric around $\omega_0$. This is a fundamental feature of the corresponding filter functions $W(\omega) = \mathcal{F} \{w (t) \}$ of these sequences in the frequency domain~\cite{Taylor08,deLange11,Naydenov11}. However, when the frequency of the magnetic field changes, we can monitor the quadrature field component and obtain phase-sensitive readout of the change. This occurs because for small changes in frequency $\delta \omega$, the field $b_z(t) = b_{ac} \cos (\omega_0 t - \delta \omega t)$, and the corresponding quadrature component $b_Q = b_{ac} \sin (\delta \omega t)$ changes linearly with $\delta \omega$. Normally as $t$ increases in the longest CP sequences, this phase change and corresponding component would be unrecoverable (see \figref{figPEA}(b)), but the excellent dynamic range of our method allows us to track the frequency as seen in \figref{figFreq}(b). SI Fig.~S6 shows simulations of the PEA when signals with equal amplitude but slightly different frequencies are used, demonstrating in principle that the PEA can distinguish such signals.

 The typical relaxation time for target nuclear spins in fluid samples at room temperature is $T_1 \sim 1 - 40$~secs~\cite{Farrar72}. The time constant $T$ for our lock-in magnetometer can be adjusted through different choices of parameters to fall within this range, as shown in the SI (Fig. S5). However, previously demonstrated technical improvements such as nano-fabricated photonic structures can greatly improve collection efficiency~\cite{Babinec10,Maletinsky12} and allow us to tune the time constant down to milleseconds, as discussed in the SI. 
 
Lastly, we discuss again the importance of linear readout and dynamic range improvements of our technique. In nanoscale magnetometry and imaging, it may be possible to have some prior knowledge on the field amplitude $b_{ac}$, from estimates of the average number of spins and the distance from the NV sensor. Recently, Refs.~\cite{Staudacher13,Mamin13} have reported breakthrough results in detecting nanoscale volumes of nuclear spins through DD noise spectroscopy with NV quantum sensors. These authors have estimated rms field amplitudes $b_{ac} \sim 70 - 400$~nT for their samples and compared the estimates with NV sensor field measurements. Ref.~\cite{Mamin13} observed significant discrepancy between the measured field compared to the estimate ($\sim 700 \%$), while Ref.~\cite{Staudacher13} used numerical modeling of the nuclear spin volumes with ``typical'' proton concentrations and other assumptions and obtained agreement at the $\sim 70 \%$ level. It is still unclear what causes the discrepancies, although the authors postulate uncertainty in either the NV position or the number of nuclear spins in the target volume leading to imperfect knowledge of the working point. While a direct comparison between our work and these results is not possible, the estimated fields are close to the maximum field amplitudes sensed in our work (see \figref{figPEA}(d)), and certainly well above that of the much longer DD sequences used by those authors. Since the sensitivity of the sequences crucially depends on this knowledge of the working point, we speculate that our methods might help in resolving some of these uncertainties. Further, since there is no restriction on the phase $\theta$ of the magnetic field, the reconstruction of the field amplitude may also have significant error if the dynamic range is limited, as we showed in \figref{figPEA}(b). Our method simultaneously resolves both the working point and phase measurement problem.

In conclusion, we have reported a new quantum sensing method for magnetometry with phase estimation algorithms. Our results show significantly improved dynamic range and linearity of the readout for time-dependent magnetic fields, while preserving the increased sensitivity of DD pulse sequences. Our method also allows for unambiguous reconstruction of the amplitude, phase, and frequency of the oscillating field, and allows us to track the phase in each measurement shot. This may open up the capability to study the spin configuration changes of nanoscale volumes of spins with unprecedented resolution, and also allow for the study of systems where the spins are hard to polarize and drive due to spectral and spatial diffusion.

\begin{acknowledgments}
This work was supported by NSF CAREER (DMR-0847195), NSF PHY-100534, DOE Early Career (DE-SC 0006638)and the Alfred P. Sloan Research Fellowship.
\end{acknowledgments}


\section*{Supplementary Information}

\setcounter{figure}{0}
\renewcommand{\thefigure}{S\arabic{figure}}

\subsection{Dynamical Decoupling Sequences: Larmor revivals}

The $^{13} C$ nuclear spin bath that has a natural abundance of $\approx 1.1 \% $ effectively produces a random field with frequency set by the nuclear gyromagnetic ratio $\gamma_n = 2 \pi (10.75)$~MHz/T and the DC bias field $B_0$. This random field causes collapses and revivals in the CP signals. For best results in AC magnetometry, it is required to operate on a revival point and this constrains the workable AC field frequencies to be $f_{ac}=\tfrac{1}{2p T_L}$  where, $p= 1,2,3 \ldots$ is an integer and $T_L$ is the Larmor period of the nuclear bath field. Having a larger bias magnetic fields could be useful for AC magnetometry due to the fast revival rates and thus giving more flexibility in terms of workable AC magnetic field frequencies.

\begin{figure}[hbt]
	\scalebox{1}[1]{\includegraphics[width=3.5in]{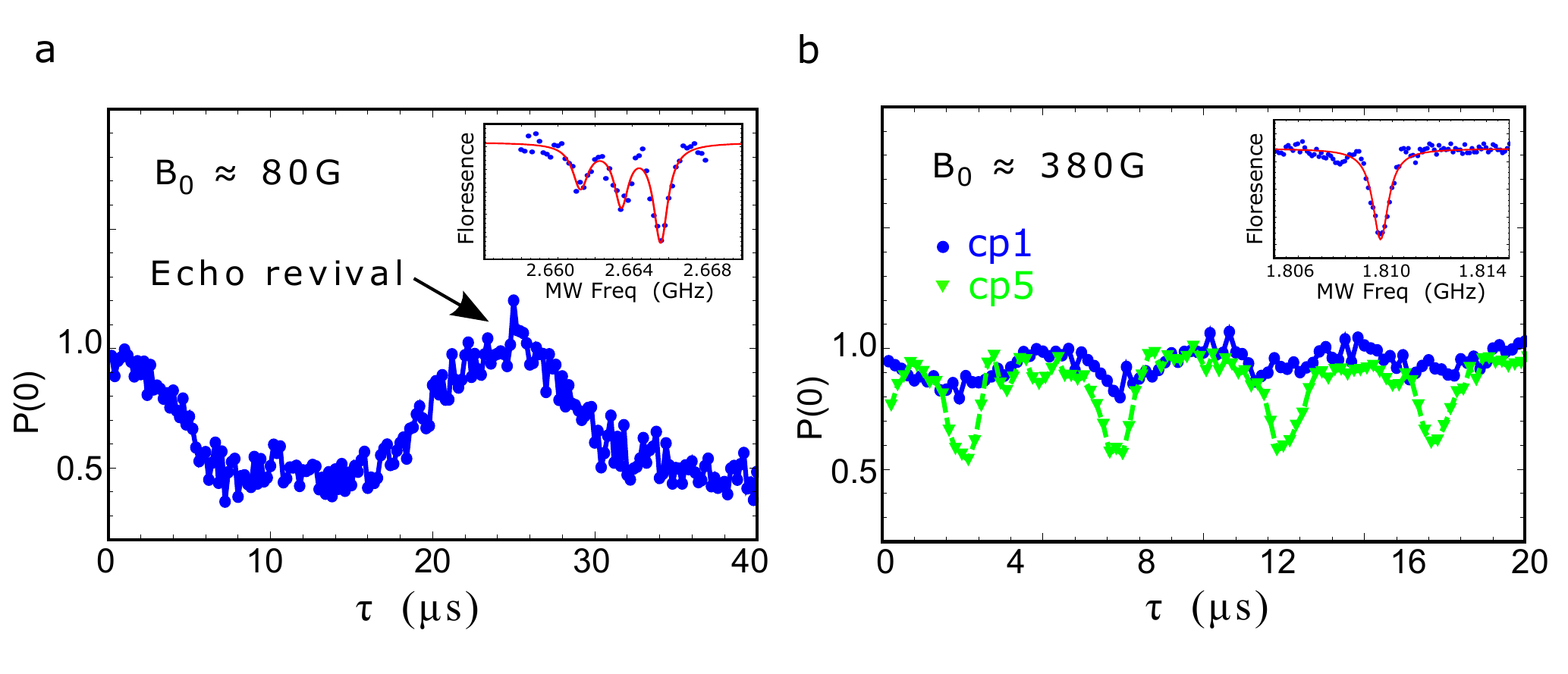}}
	\caption{\small{ Larmor revivals at \textbf{a},  $B_0 \approx 80 G $ and  \textbf{b}, $B_0 \approx 380 G $. These revivals occur due to the effective random magnetic fields arising from Larmor precession of the $^{13} C$ nuclear bath that has a natural abundance of $\approx 1.1 \% $. The insets show the corresponding ODMR spectrum for those bias fields. Magnetic fields near excited state level anti-crossing causes dynamic nuclear polarization of $^{14}$N~\cite{Jacques09}. For best results in AC magnetometry, it is required to operate on a revival point and this constrains the applicable AC field frequencies. Echo-revivals occur when $\tau = 2 p T_L$ where, $p= 1,2,3 \ldots$ is an integer and $T_L$ is the Larmor period of the nuclear bath field. Note that no external AC magnetic field was given here.}}
	\label{figRev}
\end{figure}

\subsection{Dynamical Decoupling Sequences: Coherence time enhancement}

The enhancement of the coherence time with DD sequences has been extensively studied~\cite{deLange10,Ryan10,Naydenov11,Pham12}. We use a fitting function,
\begin{equation*}
(1+D(T,T_2^{(m)}))/2 
\end{equation*}
where the $D(T,T_2^{(m)}) = \exp(- (T/T_{2}^{(m)})^\alpha)$ is the decay due to decoherence, $T_2 ^{(m)} =  T_2  m^s$, and both $\alpha$ and $s$ are sample dependent numbers which turn out to be $\alpha = 3$ and $s \approx 0.5$ in our case.

\begin{figure}[hbt]
	\scalebox{1}[1]{\includegraphics[width=3.5in]{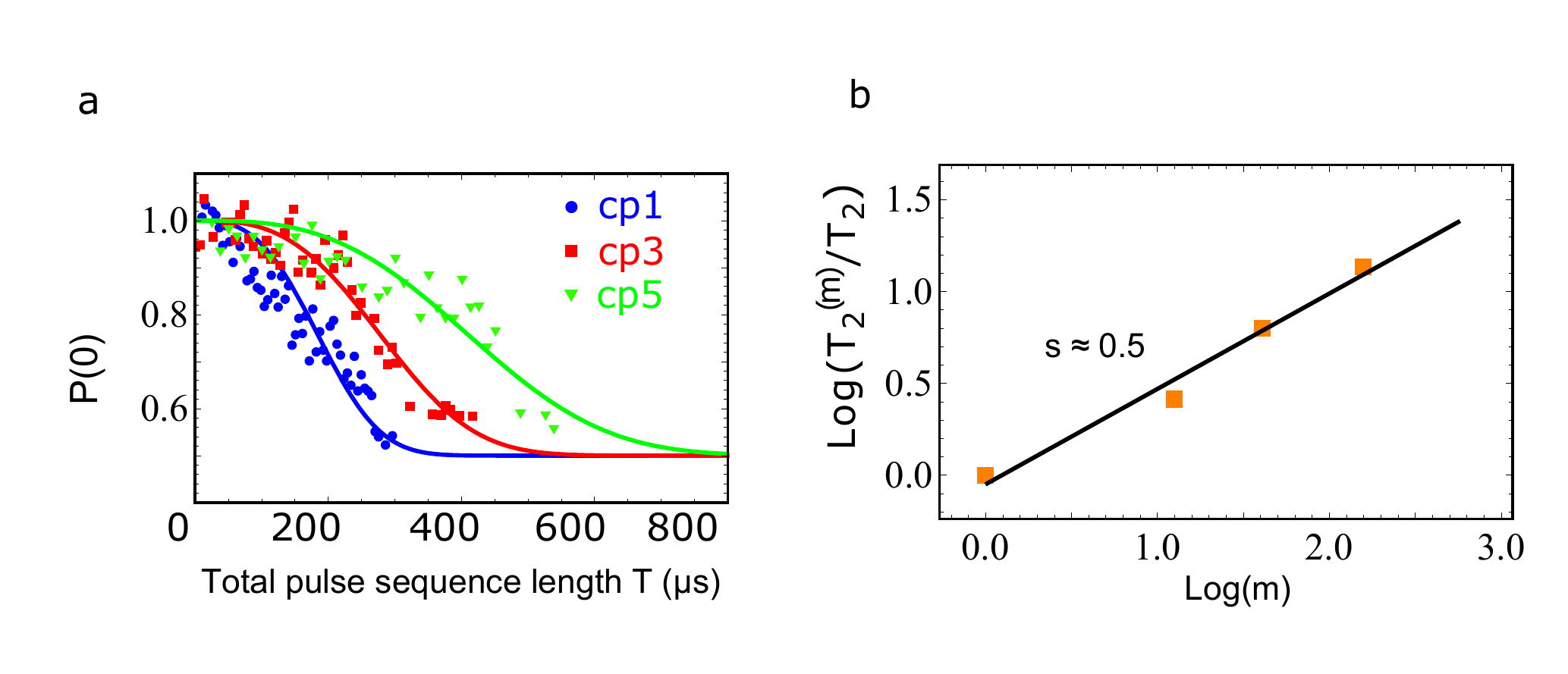}}
	\caption{\small{ \textbf{a}, Due to dynamical decoupling an enhancement in the coherence time is observed. This in turn could be employed for better sensitivity in AC magnetometry. The solid lines are the best fitted curves for the function $(1+D(T,T_2^{(m)}))/2$ where $D(T,T_2^{(m)}) = \exp(- (T/T_{2}^{(m)})^3)$ is the decay due to decoherence. \textbf{b}, The enhancement of the coherence time obeys a power law dependence: $T_2 ^{(m)} =  T_2  m^s$,  where $T_2$ is the coherence time of Hahn-Echo (CP-1), $m$ is the number of $\pi $ pulses in the CP sequence and $s$ is a sample-dependent number which turned out to be $s\approx 0.5$ in our case.}}
	\label{figCoh}
\end{figure}

\subsection{PEA: Likelihood for the unknown phase}

The probability $P(0)$ to be in $\ket{m_s=0}$ and $P(-1)$ to be in $\ket{m_s=-1}$ is related to the signal due to the in-phase magnetic field $b_I = b_{ac} \cos \theta$ by :
\begin{equation*} 
S(b_I) = 2 P(0) - 1 =  1-2 P(-1)= D(n\tau)  \cos (n \phi_I - \Phi) 
\end{equation*}
where $\phi_I = 2 \gamma_e b_I \tau/\pi $. Thus, given a quantum measurement result $u_m$  the likelihood distribution for the $\phi_I$ is given by:
\begin{equation*} 
P(u_m | \phi_I ) = \frac{\pm S(b_I)+1}{2}
\end{equation*}
where $u_m = \pm$ represents measuring $\ket{m_s=0} (\ket{m_s=-1})$ state on the $m^{th}$ measurement.
Since our aim is to find the unknown phase $\phi_I$ given the measurement results, we can use Bayes' theorem, $P(\phi_I | u_m) = P(u_m | \phi_I) P (\phi_I)/ P(u_m)$. If the \emph{a priori} distribution of the phase $P(\phi_I)$ is assumed to be flat, then $P(\phi_I | u_{m+1}) \propto P(u_m | \phi_I)$, and we multiply together the probability distributions after each measurement result followed by a normalization step to obtain the conditional probability $P_m (\phi_I)$ after all the measurements. The MLE is found from the likelihood function ($\log P_m (\phi_I)$). 

\begin{figure}[hbt]
	\scalebox{1}[1]{\includegraphics[scale=0.55]{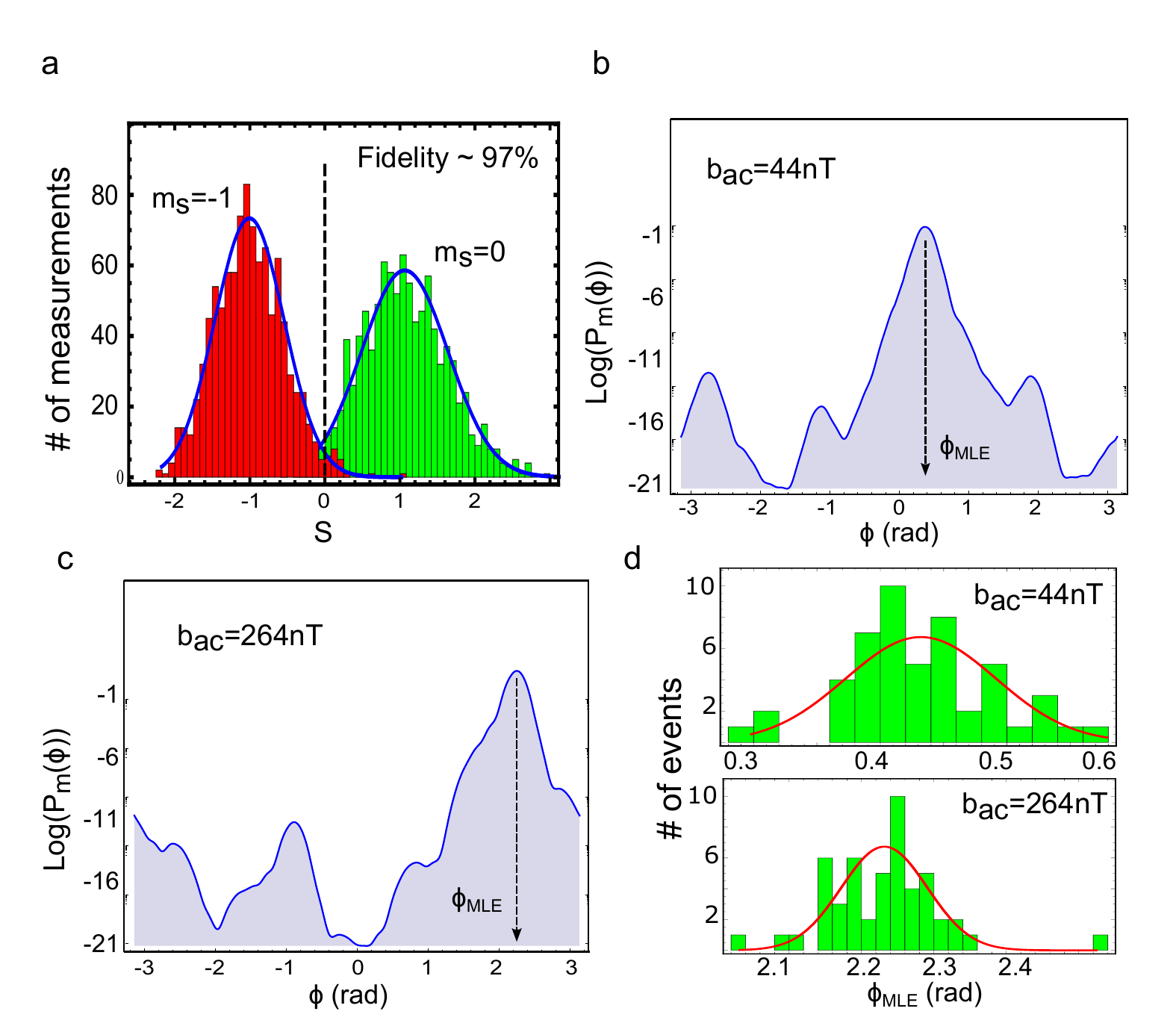}}
	\caption{\small{ \textbf{a}, A repetition of R=15000 times of the pulse sequence in our setup leads to fidelity $\sim 97 \%$ in distinguishing between the $\ket{m_s=0} $ and  $\ket{m_s=-1}$ states.  \textbf{b, c} A trial of the likelihood distribution for the unknown phase is shown for two different magnetic field amplitudes $b_{ac}:$ 44nT and 264nT respectively. The magnetic field phase and frequency are $\theta = 0^0$ and $f_{ac}=12.55$kHz. The peak of the distribution $\phi_{MLE}$ gives the maximum likelihood estimate for the unknown phase.  \textbf{d}, The histograms of  $\phi_{MLE}$'s when the experiment is repeated many (50) times for the above field amplitudes illustrates the fact that the variance of the experimental results for $\phi_{MLE}$ is more or less the same for a wide range of field amplitudes. Note that the red solid curves are Gaussian curves parametrized by the experimental results}}
	\label{figLik}
\end{figure}

\figref{figLik}(a) shows the distribution of 1000 measurement results of each state $\ket{m_s=0}$ and  $\ket{m_s=-1}$ when a single pulse sequence is repeated R=15000 times. This leads to a fidelity $\sim 97 \%$ in distinguishing the two states.  \figref{figLik}(b),(c)  shows the phase likelihood distribution for unknown quantum phase $\phi_I$ for field amplitudes $b_{ac}:$ 44nT and 264nT respectively. Here, magnetic field phase $\theta = 0^0$ and frequency $f_{ac}$=12.55kHz. \figref{figLik}(d) show the distributions of MLE's when the experiment is repeated 50 times for the above field amplitudes.

\subsection{Lockin Magnetometer: Detection of random phase jumps of any magnitude}

\begin{figure}[hbt]
	\scalebox{1}[1]{\includegraphics[scale=0.44]{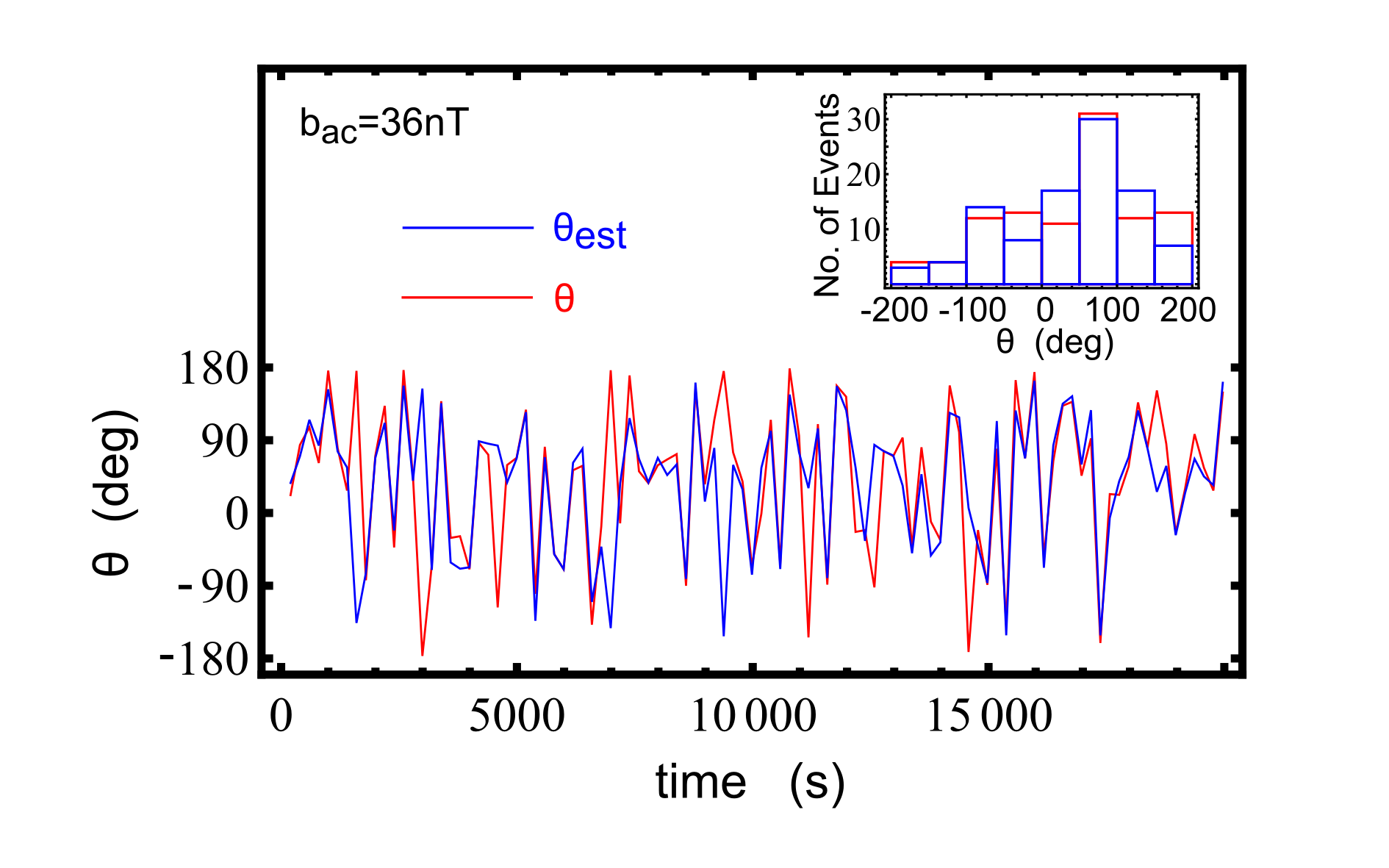}}
	\caption{\small{By monitoring both $\phi_I$ and $\phi_Q$ channels of our lock-in, we can observe any arbitrarily distributed random phase changes of the magnetic field. The time constant of the lock-in was set to $\sim 200$~sec, and the phase was held constant over this time but allowed to vary randomly between measurements. The inset histogram compares the distribution of the given external phase $\theta$ (Red) with the phase estimated  $\theta_{est}$ from the lock-in (Blue)}}
	\label{figRnd3}
\end{figure}

The PEA parameters set here were $M_K=F=4$ and leads to a time constant  $\sim 200$~sec of the lock-in, and the phase $\theta$ was held constant over this time but allowed to vary randomly between measurements. The red curve in \figref{figRnd3} shows the history of $\theta$ while the blue curve is the estimated phase $\theta_{est}$ from the lock-in. The average of the phase jumps from the estimates turned out to be $1.28^0$ with a smallest jump of $0.95^0$.

\subsection{Lockin Magnetometer: Time constants}

The relationship between the time constant of our lockin magnetometer and the parameters of the PEA are governed by the equations,
\begin{equation*}
T = \sum_{k=1}^K M(K,k) \, 2^{(k-1)} \, R \, (\tau+ t_M)
\end{equation*}

\begin{equation*}
 = R \, (\tau + t_M) [M_K (2^K -1) + F (2^K -K -1)]
\end{equation*}

where, $R$ is the number of times the pulse sequence is repeated and $t_M \sim 2 \, \mu$sec is the measurement time. 

\begin{figure}[hbt]
	\scalebox{1}[1]{\includegraphics[width=2in]{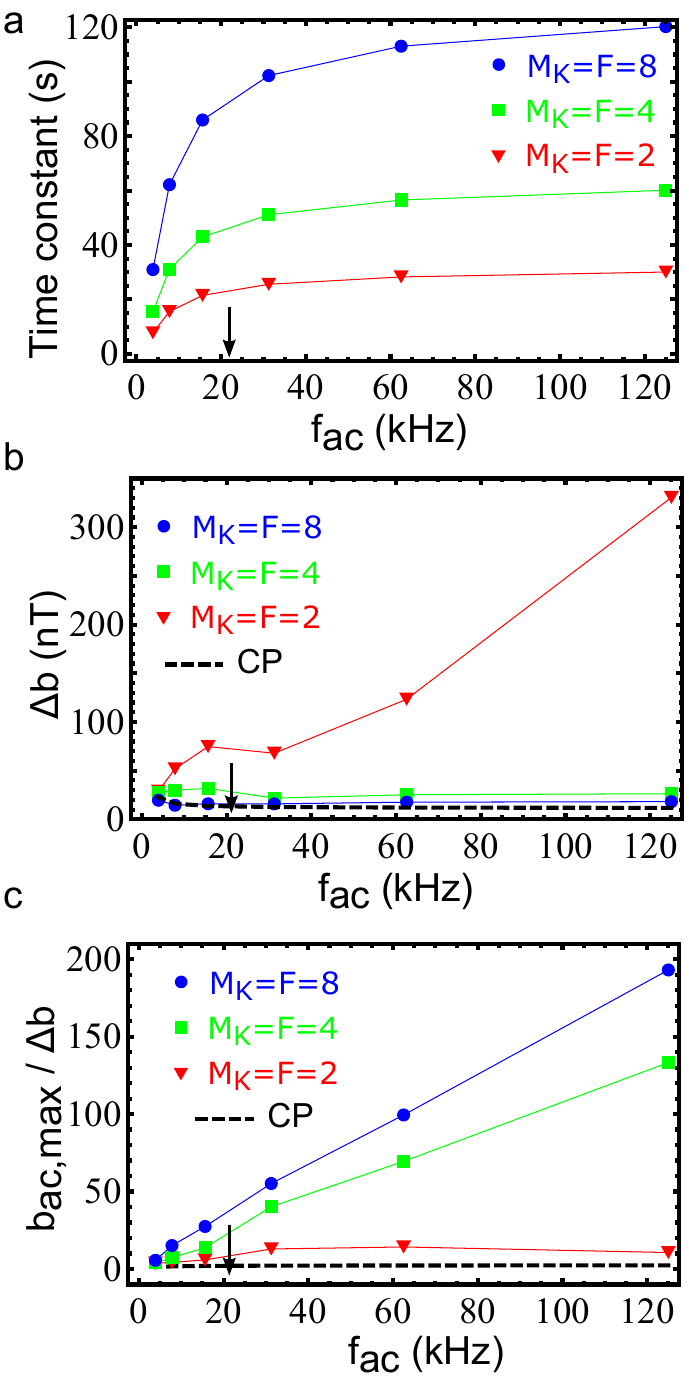}}
	\caption{\small{\textbf{PEA time constants}:  \textbf{a}, The total time for the lock-in detection as a function of magnetic field frequency $f_{ac}=1/\tau$. The parameter K is chosen such that the longest pulse sequence length is always the same ($256 \mu s$). \textbf{b}, Simulation of minimum detectable field amplitude difference $\Delta b=\pi \sigma_{\phi} /  2 \gamma_e  \tau $ as a function of $f_{ac}$ where $\sigma_{\phi}$ is the standard deviation of the simulated quantum phase readouts with measurement fidelity $\sim 93\%$ obtained from $R = 8000$  in our experiments. The black dashed line is the theoretical limit with multi-pulse CP  $\Delta b = \eta / \sqrt{T}$ for the same experimental conditions. Here, $\eta$ is obtained from Eq(3) in the main paper while the total time for PEA with $M_K=F=8$ is used for T. However note that the R=8000 is different from the actual experiments we carried out (R=15000) and also we have not taken into account the coherence enhancement due to DD in these simulations. Despite these differences, the simulation reasonably agrees with the experiments.  \textbf{c}, Corresponding dynamic range defined by $b_{ac,max}/\Delta b$.  Arrow shows the working point frequency (20.83kHz) carried out in our experiments. }}
	\label{figTimeConst2}
\end{figure}

The above equation is plotted in \figref{figTimeConst2}(a) for our experiments as a function of $f_{ac}$ for different choices of $K, M_K, F$, while keeping the longest pulse sequence length $2^{(K - 1)} \tau  = 256  \mu  $sec a constant. The longest sequence ultimately limits the sensitivity of the quantum sensing, though of course various choices of PEA parameters may result in not attaining this limit. Therefore fixing this value gives us a good way to controllable change the parameters and observe the effect on the dynamic range DR and the minimum detectable field $\Delta b$. These have to be obtained through numerical Monte-Carlo simulations that we will report in greater detail in future work, but the results are displayed in \figref{figTimeConst2}(b) and (c). As we can see there, above certain threshold measurement fidelity and choices of PEA parameters $M_K, F$, we obtain close to the decoherence limited sensitivity. By contrast, although CP can attain the decoherence limit, the DR is extremely limited.

A second factor that limits our time constant above is the factor $R$. As shown in \figref{figLik}(a), the number of repetitions of our pulse sequence will govern the fidelity with which we can make the bit measurements $u_m = \pm$. This factor $R = 1.5 \times 10^4$  in our experiments is in turn governed by the photon collection efficiency $\xi$ and visibility $V$ of the fringes in our setup. Recent improvements in these factors, e.g. through photonic nanostructures~\cite{Babinec10,Maletinsky12} or resonant excitation~\cite{Robledo11} can result in decreasing $R \rightarrow 1$, while increasing the measurement time $t_M \approx 20 \, \mu$~sec. Correspondingly our time constant can thus be reduced by almost three orders of magnitude. We also note that to integrate for longer durations, our PEA can simply be repeated more times, and the results averaged to obtain the usual improvement $\sim 1/\sqrt{T}$ in $\Delta b$.

\subsection{Lockin Magnetometer: Multiple AC frequencies}

\begin{figure}[hbt]
	\scalebox{1}[1]{\includegraphics[scale=0.55]{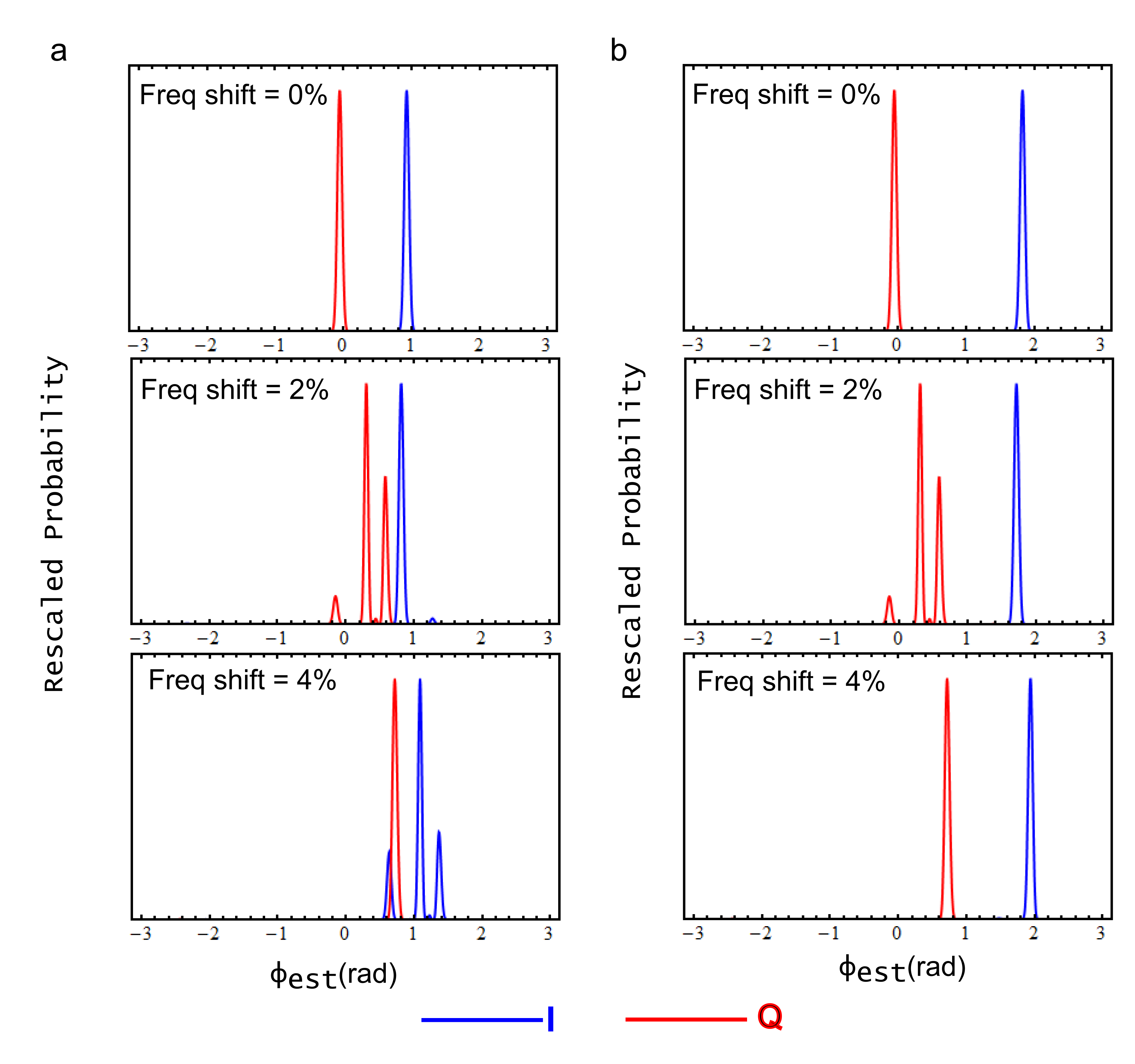}}
	\caption{\small{\textbf{Shifted frequency simulation}: \textbf{a}, The only signal is a shifted frequency relative to the lock-in frequency $f_0$=20.83kHz.  \textbf{b}, Two signals, one with a shifted frequency while the other on lock-in frequency. Blue (Red) curve plots the phase likelihood distribution obtained in I(Q) channel. All signals are set to $b_{ac}/b_{max} = 0.296$ and $\theta=0^0$.}}
	\label{fig5}
\end{figure}

Two different cases have been simulated. PEA lock-in performed on a shifted frequency relative to the lock-in frequency $f_0=20.83kHz$ is shown in \figref{fig5}(a).  The shift is given as percentage of $f_0$. The field amplitude was set to $b_{ac}/b_{max} = 0.296$  while the phase $\theta=0^0$. In this condition, the Q-channel can detect a frequencies  upto $ \sim 4 \% $ shift. \figref{fig5}(b) shows the case of two different signals one with a shifted frequency while the other on lock-in frequency given by expression: $b_{ac} (cos(2\pi f_0 t)+cos(2\pi (f_0+df_0) t))$ where $df_0$ is the frequency shift. The two in-phase signals with no frequency shift ($0\%$ shift) adds up to give a single signal with twice the amplitude. Further we can see from the figure that the $Q$ channel clearly shows the shift in the quadrature phase $\phi_Q$ that is caused by the shifted frequency component. In future work, we hope to explore further the nature and magnitude of these shifts and to develop more detailed theory and simulations.

\bibliographystyle{apsrev4-1}

%


\end{document}